\documentclass[aps,prb,twocolumn]{revtex4}
\usepackage{amsfonts}
\usepackage{amsmath}
\usepackage{graphicx}
\usepackage{dsfont}
\usepackage{diagbox}

\begin{document}

\title{Two-dimensional topological superconducting phases emerged from
d-wave superconductors in proximity to antiferromagnets}
\author{Guo-Yi Zhu$^{1}$, Ziqiang Wang$^{2}$, and Guang-Ming Zhang$^{1,3}$}
\affiliation{$^{1}$State Key Laboratory of Low-Dimensional Quantum Physics and Department
of Physics, Tsinghua University, Beijing 100084, China. \\
$^{2}$Department of Physics, Boston College, Chestnut Hill, Massachusetts
02467, USA.\\
$^{3}$Collaborative Innovation Center of Quantum Matter, Beijing 100084,
China.}
\date{\today}

\begin{abstract}
Motivated by the recent observations of nodeless superconductivity in the
monolayer CuO$_{2}$ grown on the Bi$_{2}$Sr$_{2}$CaCu$_{2}$O$_{8+\delta }$
substrates, we study the two-dimensional superconducting (SC) phases
described by the two-dimensional $t$-$J$ model in proximity to an
antiferromagnetic (AF) insulator. We found that (i) the nodal $d$-wave SC
state can be driven via a continuous transition into a nodeless $d$-wave
pairing state by the proximity induced AF field. (ii) The energetically
favorable pairing states in the strong field regime have extended $s$-wave
symmetry and can be nodal or nodeless. (iii) Between the pure $d$-wave and $%
s $-wave paired phases, there emerge two topologically distinct SC phases
with ($s+$i$d$) symmetry, i.e., the weak and strong pairing phases, and the
weak pairing phase is found to be a $Z_{2}$ topological superconductor
protected by valley symmetry, exhibiting robust gapless non-chiral edge
modes. These findings strongly suggest that the high-$T_{c}$ superconductors
in proximity to antiferromagnets can realize fully gapped symmetry protected
topological SC.
\end{abstract}

\maketitle

\section{Introduction}

Despite the intensive research in the past 30 years, the field of high-$%
T_{c} $ superconductivity (SC) in the cuprates \cite%
{Muller,Anderson,AtoZ,LeeNagaosaWen} continues to generate surprising and
challenging issues. Up to now all the high T$_{c}$ superconducting copper
oxides have layered structures, and the superconducting layers are
sandwiched by insulating charge reservoir layers. It is usually believed
that the CuO$_{2}$ layers are antiferromagnetic (AF) Mott insulators.
Modulation of charge carriers in the CuO$_{2}$ planes is realized through
substitution of chemical elements in the reservoir. In hole-doped cuprates,
it has been established that SC around the optimal doping has a $d$-wave
pairing with gap nodes along the zone diagonals\cite{Shen,Harlingen,Tsuei}.

\begin{figure}[t]
\includegraphics[width=8cm]{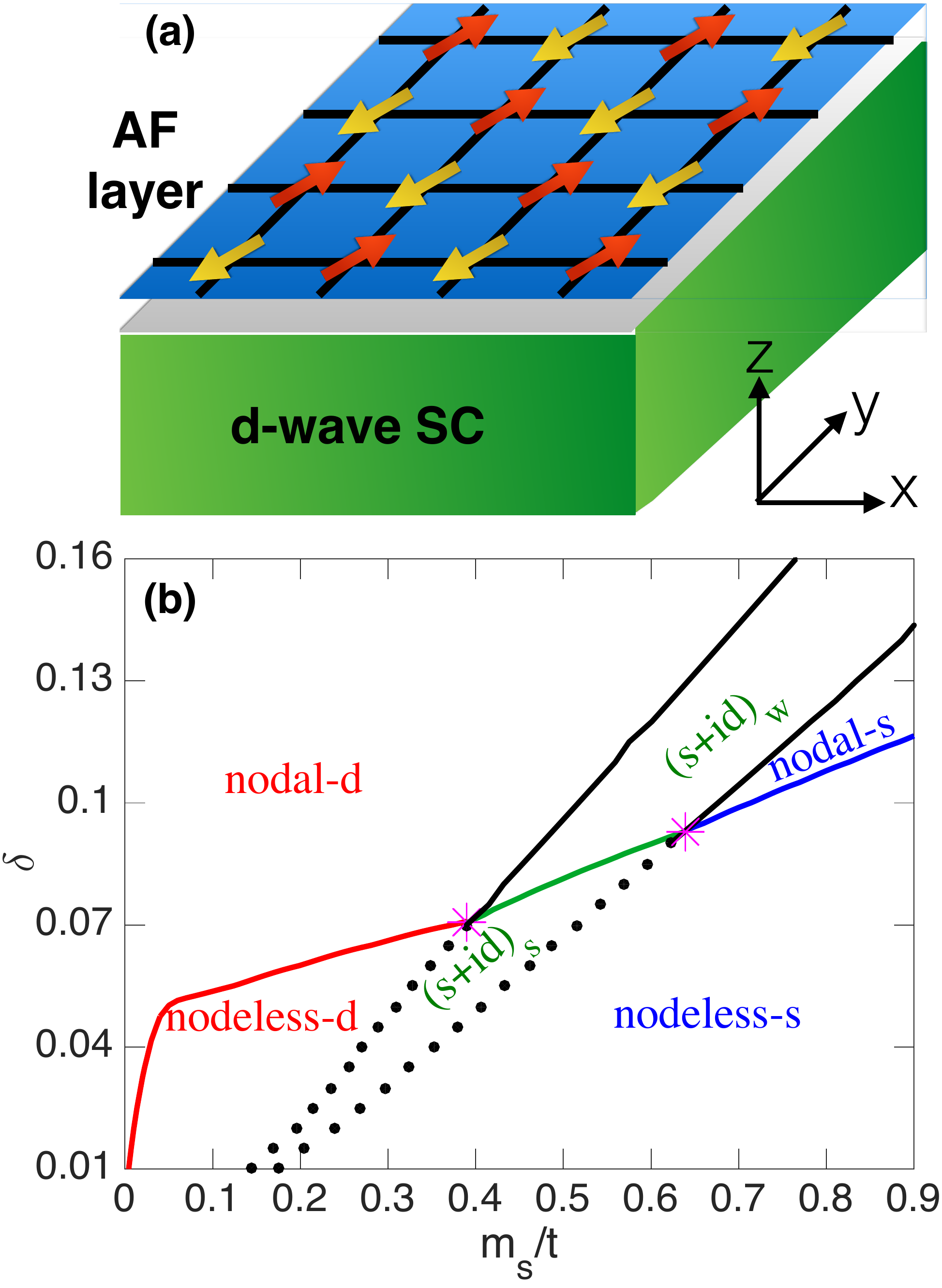} \vskip-0.5cm
\caption{(a) An antiferromagnetic insulating CuO$_{2}$ layer is grown on the
optimal doped Bi-2212 copper oxide substrates. (b) The phase diagram of
superconducting phases in terms of the AF field $m_{s}$ and doping $\protect%
\delta $ is derived with the relevant parameters for the cuprates: $J/t=0.3$%
, $t^{\prime }/t=0.2$ and $0.01<\protect\delta <0.16$. Continuous phase
transitions are marked by solid lines, whereas discontinuous phase
boundaries by dotted lines. Red, green and blue lines separate the weak
pairing phases (upper region) from the strong pairing phases (lower region).}
\label{setup&phase_diagram}
\end{figure}

Recently, Zhong, et. al. reported that a monolayer CuO$_{{2}}$ is
successfully grown {on the optimal doped Bi}$_{{2}}${Sr}$_{{2}}${CaCu}$_{{2}%
} ${O}$_{{8+}\delta }${{\ (Bi-2212) substrates via molecular beam epitaxy%
\cite{Xue}}, enabling direct probe of the CuO$_{2}$ plane by scanning
tunneling microscopy. Their results are interesting and important. Unlike
the sandwiched CuO}$_{2}${\ layers in the bulk Bi-2212, the overall
electronic spectral density on the monolayer films is characterized by a
large (}$\sim $ {$2$ eV) Mott-Hubbard-like gap\cite{Xue}. In the low-energy
regime, however, two distinct and spatially separated energy gaps are
observed on the films: the V-shaped gap is similar to the gap observed on
the BiO layer, and the U-shaped gap }is identified with the superconducting
nature\cite{Xue}. Such an U-shaped superconducting gap is in striking
contrast with the nodal gap in the d$_{x^{2}-y^{2}}$-wave pairing symmetry%
\cite{Shen,Harlingen,Tsuei}. Therefore, the observed superconductivity in
the monolayer CuO$_{2}$ on optimal doped Bi-2212 substrates raises a
challenge whether a new nodeless SC appears at the interface between nearly
optimal doped Bi-2212 and the {CuO}$_{{2}}$ AF insulating layer (see Fig.~%
\ref{setup&phase_diagram}a).

Motivated by this recent experiment, we examine the possible superconducting
phases derived from standard $t$-$J$ model with nearest neighbor singlet
pairings in proximity to an antiferromagnetic insulating layer. In the
absence of the AF layer, the superconducting state of the t-J model has a
purely d-wave pairing phase with four nodes for doping up to $\delta =0.25$.
The proximity effect induced by the AF layer is modeled by applying an
external AF field. With the parameters being relevant for the cuprates: $%
J/t=0.3$, $t^{\prime }/t=0.2$ and $0.01<\delta <0.16$, our findings are
summarized in the phase diagram in terms of hole doping $\delta $ and the AF
field $m_{s}$, shown in Fig.~\ref{setup&phase_diagram}b. From weak to
moderate of $m_{s}$, we found a continuous transition from the nodal $d$%
-wave to a nodeless $d$-wave SC with reduced doping or increased $m_{s}$.
However, the strong AF field drives the $d$-wave pairing unstable and
extended $s$-wave pairing is more favored energetically, leading to a fully
gapped $s$-wave phase and a continuous transition to nodal $s$-wave pairing
at higher doping.

In the intermediate region of $m_{s}$, the pure $d$-wave and $s$-wave
pairings are degenerate in energy such that the SC states with mixed ($s+id$%
) pairing emerge as the energetically most favorite. These mixed pairing
states preserve valley and mirror symmetries. Remarkably, we found that
there exist two topologically distinct gapped SC states, termed as $%
(s+id)_{s}$ and $(s+id)_{w}$ corresponding to the \emph{strong} and \emph{%
weak} pairing SC phases, respectively\cite{ReadGreen}. The $(s+id)_{w}$
phase is identified as a $Z_{2}$ topological SC protected by the valley
symmetry and supports robust gapless non-chiral edge modes. Our findings
strongly suggest that the high-$T_{c}$ copper-oxide superconductors in
proximity to AF insulating phases can not only produce fully gapped SC
states of various pairing symmetries but also potentially realize
topological valley SC.

\section{Model and theory}

We start to consider the square lattice $t$-$J$ model for the d-wave
superconductor in proximity to an AF insulating layer (Fig.~\ref%
{setup&phase_diagram}a),
\begin{eqnarray}
H &=&-t\sum_{\mathbf{r},\mathbf{\eta },\sigma }c_{\mathbf{r},\sigma
}^{\dagger }c_{\mathbf{r+\eta },\sigma }+t^{\prime }\sum_{\mathbf{r,\gamma }%
,\sigma }c_{\mathbf{r},\sigma }^{\dagger }c_{\mathbf{r+\gamma },\sigma }
\notag \\
&&+\frac{J}{2}\sum_{\mathbf{r,\eta }}\left( S_{\mathbf{r}}\cdot S_{\mathbf{%
r+\eta }}-\frac{1}{4}n_{\mathbf{r}}n_{\mathbf{r+\eta }}\right)   \label{h} \\
&&-\mu _{0}\sum_{\mathbf{r,\sigma }}c_{\mathbf{r},\sigma }^{\dagger }c_{%
\mathbf{r},\sigma }+m_{s}\sum_{\mathbf{r,\sigma }}\sigma (-1)^{\mathbf{r}}c_{%
\mathbf{r},\sigma }^{\dagger }c_{\mathbf{r},\sigma },  \notag
\end{eqnarray}%
where $t$ and $t^{\prime }$ are the nearest neighbor (NN) and next nearest
neighbor (NN) hoppings, $\eta $ and $\gamma $ denote the corresponding
vectors, $J$ is the NN Heisenberg exchange interaction, and the proximate AF
insulating layer has an in-plane staggered field $m_{s}$ along the $y$
direction. The square lattice is bipartitioned into a checkerboard A-B
sublattice. Although the AF ordering inevitably breaks the time reversal
symmetry $\mathcal{T}=i\sigma _{y}\mathcal{K}$, each unit cell shows zero
net magnetic field, and the product of the time reversal and one unit
lattice translation $\tau _{z}$ (equivalent to switching A-B sub-lattices) $%
\mathcal{\tilde{T}}\equiv i\sigma _{y}\tau _{z}\mathcal{K}$ is respected,
where $\tau _{z}$ denotes the Pauli matrix acting upon the sublattice spinor
space with the eigen-spinor of $\tau _{x}=\pm 1$ living on A/B sublattice.
The site-centered mirror symmetry with respect to the $y=0$ plane $%
M_{y}=i\sigma _{y}$ is respected, while the mirror reflection regarding $x=0$
or $z=0$ plane is preserved only by combining unit lattice translation: $%
\tilde{M}_{x/z}\equiv i\sigma _{x/z}\tau _{z}$, which is the bond-centered
mirror reflection symmetry. Thanks to the singlet pairing nature, shifting
the AF field from in-plane to out-of-plane only adapts the mirror symmetries
analysis, while most of our results still hold.

In this paper, we fix the parameters to be relevant for the cuprates: $%
J/t=0.3$, $t^{\prime }/t=0.2$ and $0.01<\delta <0.16$. The local constraint
of no double occupancy $\sum_{\sigma }c_{r,\sigma }^{\dagger }c_{r,\sigma
}\leq 1$ has to be imposed. Writing the electron operators in terms of a
fermionic spinon and a bosonic holon: $c_{r,\sigma }=b_{r}^{\dagger
}f_{r,\sigma }$, the constraint changes into an equality $b_{r}^{\dagger
}b_{r}+\sum_{\sigma }f_{r,\sigma }^{\dagger }f_{r,\sigma }=1$, which can be
enforced using a Lagrangian multiplier $\lambda $. The doping concentration
is given by $\delta =\langle b_{r}^{\dagger }b_{r}\rangle $ and we have $%
\sum_{\sigma }\langle f_{r,\sigma }^{\dagger }f_{r,\sigma }\rangle =1-\delta
$.

In the mean-field (MF) approach to the SC state\cite%
{KotliarLiu,Joynt,LeeNagaosaWen}, the holons condense, i.e., $b_{r}^{\dagger
}$ and $b_{r}$ are replaced by their expectation value $\sqrt{\delta }$. The
superexchange interaction in Eq.(\ref{h}) can be decoupled in the
paramagnetic valence bond and spin-singlet pairing channels by introducing
the order parameters
\begin{eqnarray}
\chi  &\equiv &\frac{J}{4}\langle f_{r,\uparrow }^{\dagger }f_{r+\eta
,\uparrow }+f_{r,\downarrow }^{\dagger }f_{r+\eta ,\downarrow }\rangle ,
\notag \\
\Delta _{\eta } &\equiv &\frac{J}{4}\langle f_{r,\uparrow }f_{r+\eta
,\downarrow }-f_{r,\downarrow }f_{r+\eta ,\uparrow }\rangle .
\end{eqnarray}%
In general, we assume $\Delta _{x}=\Delta _{s}+i\Delta _{d}$ and $\Delta
_{y}=\Delta _{s}-i\Delta _{d}$, where $\Delta _{s}$ and $\Delta _{d}$ are
amplitudes of NN spin-singlet pairing with $s_{x^{2}+y^{2}}$- and $%
d_{x^{2}-y^{2}}$-symmetries, respectively. In momentum space, the MF
Hamiltonian can be written in terms of Nambu spinors $F_{\mathbf{k}%
}^{\dagger }=(f_{\mathbf{k}\uparrow }^{\dagger },f_{\mathbf{k+Q}\uparrow
}^{\dagger },f_{-\mathbf{k}\downarrow },f_{-\mathbf{k-Q}\downarrow })$,
where $\mathbf{Q}=(\pi ,\pi )$ is the AF wave vector. Writing $H_{MF}=\frac{1%
}{2}\sum_{\mathbf{k}}F_{\mathbf{k}}^{\dagger }H_{\mathbf{k}}F_{\mathbf{k}}$
in unfolded Brillouin zone, we derive the Hamiltonian matrix as
\begin{equation*}
H_{\mathbf{k}}=\left(
\begin{array}{cc}
m_{s}\tau ^{x}+\epsilon _{\mathbf{k}}\tau ^{z}+\epsilon _{\mathbf{k}%
}^{\prime }-\mu  & \Delta _{\mathbf{k}}\tau ^{z} \\
\Delta _{\mathbf{k}}^{\ast }\tau ^{z} & m_{s}\tau ^{x}-\epsilon _{\mathbf{k}%
}\tau ^{z}-\epsilon _{\mathbf{k}}^{\prime }+\mu
\end{array}%
\right) ,
\end{equation*}%
where $\mu \equiv \mu _{0}-\lambda $ is the renormalized chemical potential,
$\epsilon _{\mathbf{k}}=-2(t\delta +\chi )\left( \cos k_{x}+\cos
k_{y}\right) $, $\epsilon _{\mathbf{k}}^{\prime }=4t^{\prime }\delta \cos
k_{x}\cos k_{y}$, and the SC gap function is defined by%
\begin{equation}
\Delta _{\mathbf{k}}=2\Delta _{d}\left( \cos k_{x}-\cos k_{y}\right)
-i2\Delta _{s}\left( \cos k_{x}+\cos k_{y}\right)
\end{equation}%
up to a global phase that can be gauge-fixed.

It should be emphasized that the MF Hamiltonian with pure singlet pairing
does not break the symmetry $\mathcal{\tilde{T}}$ and mirror symmetries of
the prototypical Hamiltonian, but the mixed phase of $s$-wave and $d$-wave
pairings would break the symmetry $\mathcal{\tilde{T}}$ spontaneously. To
make the physics more transparent, we adopt a two-step strategy to
diagonalize the MF Hamiltonian: first the normal sector spanned by AF
ordering is diagonalized as:
\begin{equation}
\xi _{\pm ,\mathbf{k}}=\epsilon _{\mathbf{k}}^{\prime }-\mu \pm \sqrt{%
\epsilon _{\mathbf{k}}^{2}+m_{s}^{2}},  \label{quasiSpc}
\end{equation}%
with corresponding AF quasiparticles%
\begin{eqnarray}
\psi _{+,\mathbf{k},\sigma }^{\dagger } &=&\left( \text{cos}\theta _{\mathbf{%
k}}\right) f_{\mathbf{k},\sigma }^{\dagger }+\sigma \left( \text{sin}\theta
_{\mathbf{k}}\right) f_{\mathbf{Q}+\mathbf{k},\sigma }^{\dagger },  \notag \\
\psi _{-,\mathbf{k},\sigma }^{\dagger } &=&\left( \text{sin}\theta _{\mathbf{%
k}}\right) f_{\mathbf{k},\sigma }^{\dagger }-\sigma \left( \text{cos}\theta
_{\mathbf{k}}\right) f_{\mathbf{Q}+\mathbf{k},\sigma }^{\dagger },
\label{slave2quasi}
\end{eqnarray}%
where $\theta _{\mathbf{k}}\equiv \frac{1}{2}\tan ^{-1}\frac{m_{s}}{\epsilon
_{\mathbf{k}}}\in \left[ 0,\frac{\pi }{2}\right] $. The normal state owns a
two-fold Kramer's degeneracy due to the symmetry $\tilde{\mathcal{T}}$,
which is anti-unitary and $\tilde{\mathcal{T}}^{2}=-1$. Furthermore, the
symmetry $\mathcal{\tilde{T}}$ acting on the slave spinor is equivalent to
the time-reversal $\mathcal{T}$ acting on the AF quasiparticles, and
similarly for the mirror symmetries:
\begin{eqnarray}
\mathcal{\tilde{T}}^{-1}\left(
\begin{array}{c}
f_{\mathbf{k},\sigma } \\
f_{\mathbf{k}+\mathbf{Q},\sigma }%
\end{array}%
\right) \tilde{\mathcal{T}}\text{ } &\Longleftrightarrow &\mathcal{T}%
^{-1}\left(
\begin{array}{c}
\psi _{+,\mathbf{k},\sigma } \\
\psi _{-,\mathbf{k},\sigma }%
\end{array}%
\right) \mathcal{T},  \notag \\
\tilde{M}_{x/z}^{-1}\left(
\begin{array}{c}
f_{\mathbf{k},\sigma } \\
f_{\mathbf{k}+\mathbf{Q},\sigma }%
\end{array}%
\right) \tilde{M}_{x/z}\text{ } &\Longleftrightarrow &M_{x/z}^{-1}\left(
\begin{array}{c}
\psi _{+,\mathbf{k},\sigma } \\
\psi _{-,\mathbf{k},\sigma }%
\end{array}%
\right) M_{x/z},  \notag \\
M_{y}^{-1}\left(
\begin{array}{c}
f_{\mathbf{k},\sigma } \\
f_{\mathbf{k}+\mathbf{Q},\sigma }%
\end{array}%
\right) M_{y}\text{ } &\Longleftrightarrow &M_{y}^{-1}\left(
\begin{array}{c}
\psi _{+,\mathbf{k},\sigma } \\
\psi _{-,\mathbf{k},\sigma }%
\end{array}%
\right) M_{y}.
\end{eqnarray}%
Physically, this suggests the original time-reversal $\mathcal{T}$ and $%
M_{x/z}$ restore for the AF quasi-particles without combining the sublattice
transformations, which can be understood as that AF quasi-particles simply
lose sight of the sublattice. The inter-band pairing between these two
species of AF quasi-particles are found to be absent due to the NN singlet
pairing nature. Hence the MF Hamiltonian is decoupled into: $H_{\text{MF}}=%
\frac{1}{4}\sum_{\mathbf{k},\alpha =\pm }\Psi _{\alpha ,\mathbf{k}}^{\dagger
}H_{\alpha }(\mathbf{k})\Psi _{\alpha ,\mathbf{k}}$, where
\begin{equation}
H_{\pm }(\mathbf{k})=\xi _{\pm ,\mathbf{k}}\rho _{z}\pm (\Delta _{\mathbf{k}%
}\rho _{+}+\Delta _{\mathbf{k}}^{\ast }\rho _{-}),
\end{equation}%
with the Nambu spinor $\Psi _{\pm ,\mathbf{k}}^{\dagger }\equiv \left(
\begin{array}{cccc}
\psi _{\pm ,\mathbf{k},\uparrow }^{\dagger } & \psi _{\pm ,\mathbf{k}%
,\downarrow }^{\dagger } & \psi _{\pm ,-\mathbf{k},\downarrow } & -\psi
_{\pm ,-\mathbf{k},\uparrow }%
\end{array}%
\right) $. Thus, the energy spectrum appears pairwise as $\pm E_{\pm }(%
\mathbf{k})$,
\begin{equation}
E_{\pm }\left( \mathbf{k}\right) =\sqrt{\xi _{\pm }^{2}\left( \mathbf{k}%
\right) +|\Delta _{\mathbf{k}}|^{2}}.
\end{equation}%
The SC ground-state energy density is given by
\begin{eqnarray}
\varepsilon _{g} &=&-\frac{1}{2N}\sum_{\mathbf{k}}[E_{+}\left( \mathbf{k}%
\right) +E_{-}(\mathbf{k})]+\frac{J}{8}\delta (1-\delta )-\mu \delta  \notag
\\
&&+\frac{8}{J}\left( \kappa ^{2}+\Delta _{s}^{2}+\Delta _{d}^{2}\right) +\mu
_{0}(\delta -1).
\end{eqnarray}%
Then the saddle point equations are obtained by minimizing the ground state
energy $\partial \varepsilon _{g}/\partial (\chi ,\Delta _{s},\Delta
_{d},\mu )=0$, from which the MF parameters $(\chi ,\Delta _{s},\Delta
_{d},\mu )$ are determined self-consistently.

\section{Phase diagram of superconducting phases}

\begin{figure}[t]
\includegraphics[width=8cm]{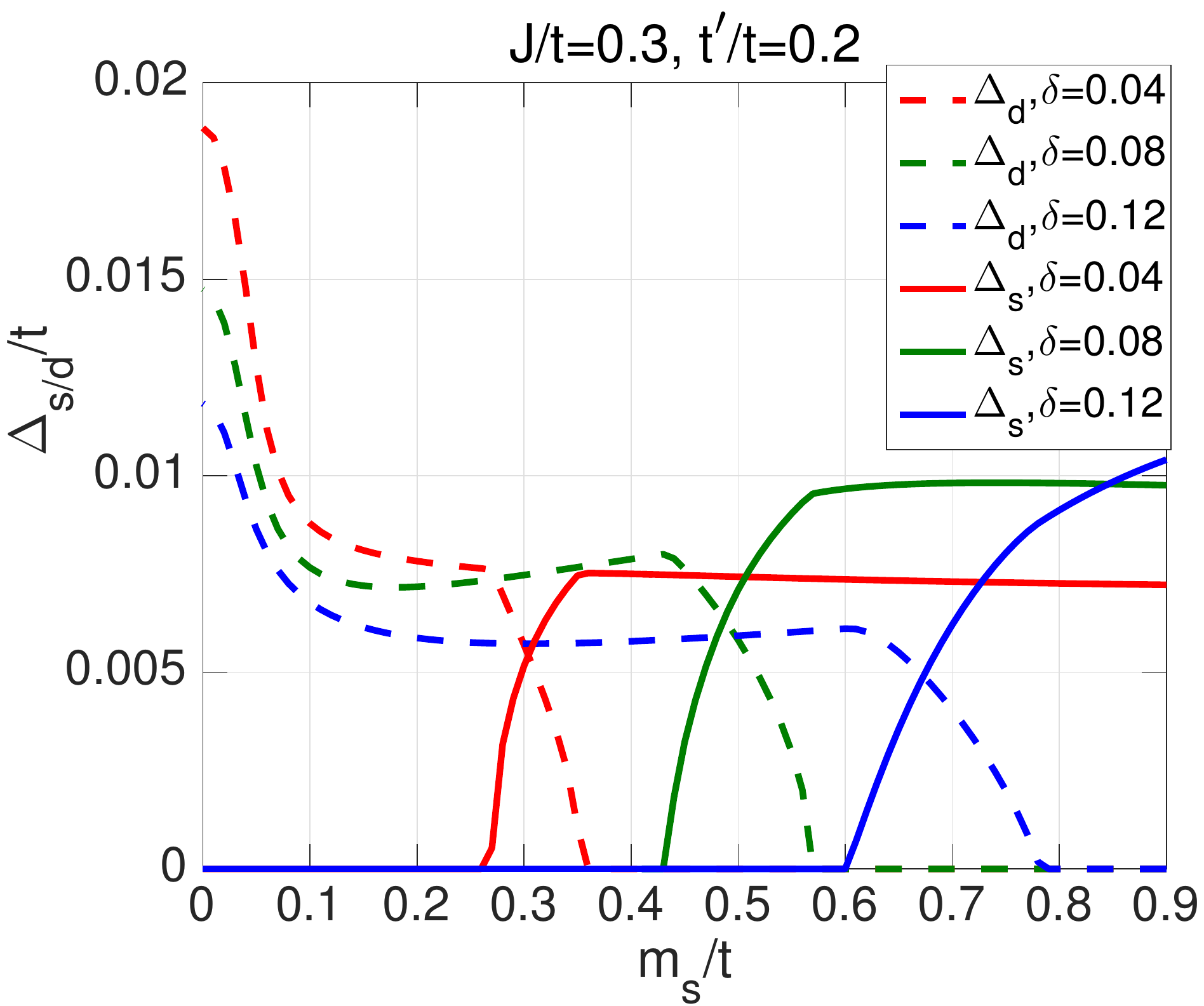}
\caption{Pairing amplitudes of SC quasiparticles. The $d_{x^{2}-y^{2}}$-wave
(dotted lines) and extended $s_{x^{2}+y^{2}}$-wave (solid lines) pairing
amplitudes as a function of $m_{s}$ for a given doping $\protect\delta $.}
\label{pairing_order}
\end{figure}

The complete phase diagram in the $m_{s}$-$\delta $ plane (Fig.~\ref%
{setup&phase_diagram}b) is obtained by minimizing the ground state energy
and considering the nodeness of quasiparticle spectrum. As is shown in Fig.~%
\ref{pairing_order}, when AF field $m_{s}$ is relatively weak, $%
d_{x^{2}-y^{2}}$-wave pairing symmetry dominates which concurs with the
consensus. However, a strong proximate AF field drives the energetically
favorite pairing symmetry from $d_{x^{2}-y^{2}}$-wave to $s_{x^{2}+y^{2}}$%
-wave continuously through a mixed pairing regime $%
s_{x^{2}+y^{2}}+id_{x^{2}-y^{2}}$. Further, each region is bisected into two
different phases.

In the $d_{x^{2}-y^{2}}$-wave phase, the nodes of the lower Bogoliubov
quasi-particle spectrum $E_{-}(\mathbf{k})$ appear when the zero lines of $%
\xi _{-}(\mathbf{k})=0$ intercepts that of $\Delta _{\mathbf{k}}=0$, i.e.
satisfying the condition $a\cos ^{2}2k_{x}+b\cos 2k_{x}+c=0$, where $%
a=(t^{\prime }\delta )^{2}$, $b=2(t^{\prime }\delta )^{2}-\mu t^{\prime
}\delta -2(t\delta +\chi )^{2}$, and $c=\left( t^{\prime }\delta -\mu
/2\right) ^{2}-2(t\delta +\chi )^{2}-m_{s}^{2}/4$. According to this
criterion, the $d_{x^{2}-y^{2}}$-wave region is divided into the nodal phase
for moderate doping or a weak AF field, and nodeless phase for underdoping
or relatively strong AF field. Quite similarly, the $s_{x^{2}+y^{2}}$-wave
phase shows nodes in spectrum when $\cos 2k_{x}=-\frac{m_{s}+\mu }{%
2t^{\prime }\delta }-1$ is satisfied, and is therefore divided into nodal
and nodeless phases as well. The mixed pairing $%
s_{x^{2}+y^{2}}+id_{x^{2}-y^{2}}$ also consists of two fully gapped phases $%
(s+id)_{w}$ and $(s+id)_{s}$ separated by a critical line on which the
spectrum exhibits four nodes at $X_{\pm }\equiv (\pi /2,\pm \pi /2)$ and
their inversion partners.

In Fig.~\ref{setup&phase_diagram}b, the critical line of the nodal and
nodeless $d_{x^{2}-y^{2}}$-wave SC (marked by red line) connects the
critical line dividing the $(s+id)_{w}$ and $(s+id)_{s}$ phases (green
line), which further joins the critical line separating the nodal and
nodeless $s_{x^{2}+y^{2}}$-wave phases (blue line). This is guaranteed by
the hidden topological nature. In fact, as we shall show later, the nodal $%
d_{x^{2}-y^{2}}$, $(s+id)_{w}$ and nodal $s_{x^{2}+y^{2}}$ SC phases can be
classified as the weak pairing and topologically nontrivial, while the
nodeless-$d_{x^{2}-y^{2}}$, $(s+id)_{s}$ and nodeless-$s_{x^{2}+y^{2}}$ as
the strong pairing and topologically trivial. So the joint critical lines
penetrating the phase diagram are essentially the phase transition from weak
pairing to strong pairing regardless of pairing symmetry.

\section{Nodal $d$-wave SC and its phase transitions}

The nodal $d$-wave SC in the presence of AF field shows a pair of
inequivalent nodes in the first quadrant of the unfolded Brillouin zone as a
result of the band folding. The locations of the nodes are denoted by $%
\mathbf{K}_{\pm }\equiv \left( K_{\pm },K_{\pm }\right) $ ($0<K_{+}<\pi /2$
and $K_{-}=\pi -K_{+}>\pi /2$). The nodes located on the other quadrants of
the Brillouin zone are related to $\mathbf{K}_{\pm }$ by mirror reflection
or inversion. The underlying topology is encoded in the low-energy
Bogoliubov quasi-particles in the vicinity of these nodes. To examine the
low-energy effective Hamiltonian, we expand the MF Hamiltonian around the
nodal points $\mathbf{K}_{\pm }$. In the AF quasiparticle basis, $\Psi
_{+}^{\dagger }(\mathbf{k})$ is fully gapped and frozen in low-energy limit,
therefore, the BdG effective Hamiltonian for $\Psi _{-}^{\dagger }(\mathbf{k}%
)$ is obtained
\begin{equation}
H_{\text{eff}}\left( \mathbf{K}_{\pm }+\mathbf{q}\right) =\pm v_{3}q_{+}\rho
_{z}+v_{1}q_{-}\rho _{x}\equiv \vec{h}_{\pm }(\mathbf{q})\cdot \vec{\rho},
\label{NodalEffHam}
\end{equation}%
where $q_{\pm }\equiv q_{x}\pm q_{y}$ and two characteristic velocities: $%
v_{1}=2\Delta _{d}\sin K_{+}$ and $v_{3}=-2t^{\prime }\delta \sin
2K_{+}+4(t\delta +\kappa )^{2}\sin 2K_{+}/\sqrt{m_{s}^{2}+16(t\delta +\kappa
)^{2}\cos ^{2}K_{+}}$. Actually this matrix is similar to that describing a
pair of two-dimensional Weyl fermions with opposite chirality around $%
\mathbf{K}_{\pm }$, and the pseudo-magnetic field $\vec{h}_{\pm }(\mathbf{k})
$ exhibits anti-vortex/vortex topological texture with nodes being the
vortex cores. This underlying topology entails robust Andreev bound states
on the edges with momentum residing between the projection of nodes\cite%
{ZhuZhang}. For convenience, Eq.(\ref{NodalEffHam}) only shows the two
valleys in the first quadrant of Brillouin zone while leaving their mirror
partners on $\pm (K_{\pm },-K_{\pm })$ and time reversal partners on $-%
\mathbf{K}_{\pm }$ behind. The whole system preserves the mirror symmetries
and emergent $\mathcal{T}$ symmetry. In fact, the nontrivial topology of the
nodal $d$-wave SC is protected by emergent $\mathcal{T}$ symmetry, which
forbids the mass term proportional to the matrix $\rho _{y}$ and confines $%
\vec{h}_{\pm }(\mathbf{q})$ to lie in-plane. So the anti-vortex/vortex
structure is guaranteed and cannot be destroyed by arbitrary weak
perturbations.

Actually there are two distinct ways to gap out the nodes by bestowing a
mass term upon the Weyl fermion-like quasiparticles. The first one is to
gradually tune the positions of a pair of nodes to merge so that the
coupling of quasiparticles generates a mass for each other, driving them
into massive Dirac fermion. This is the only way allowed by the emergent $%
\mathcal{T}$ symmetry. From the perspective of $\vec{h}_{\pm }(\mathbf{q})$,
the vortex and anti-vortex annihilate with each other, killing the
nontrivial topology. This process can be achieved by increasing the AF field
or decreasing the doping concentration, and the weak pairing nodal $d$-wave
SC thus changes into the nodeless $d$-wave SC through a continuous phase
transition\cite{Kivelson,Yuanming}. Evolution of the Bogoliubov
quasiparticle spectrum through this transition is shown in Fig.~\ref%
{WeakPairingDwavePhaseTransitionSpcEvol}. The critical point is
characterized by a highly anisotropic Bogoliubov dispersion: along the nodal
line the dispersion is quadratic nonrelativistic while perpendicular to the
nodal line it is the linear Dirac dispersion.
\begin{figure}[t]
\includegraphics[width=8cm]{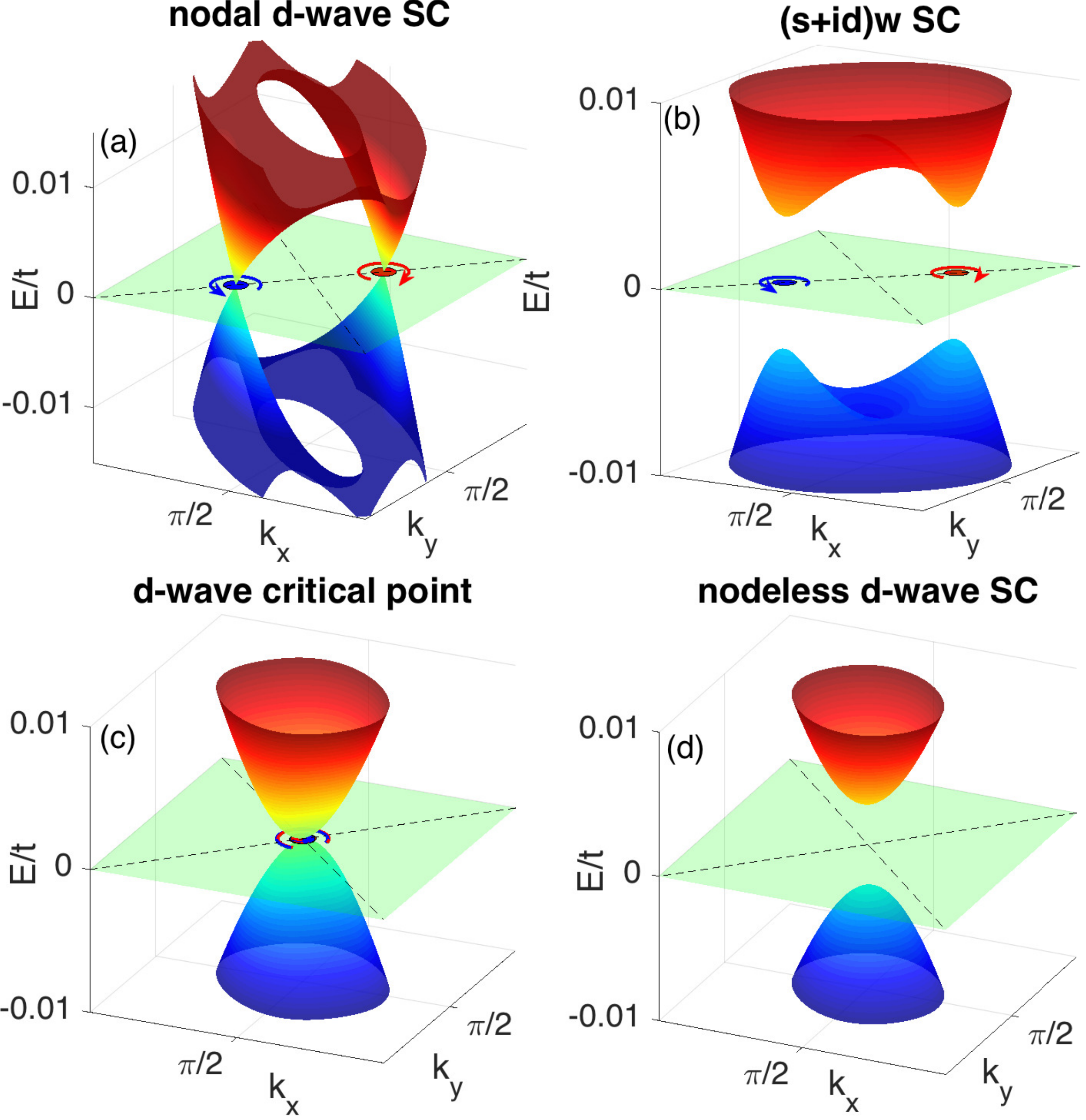} %
\vskip-0.5cm
\caption{Low-energy Bogoliubov quasi-particle energy dispersion of various
SC phases in the first quadrant of the unfolded Brillouine zone. The blue
and red cycling arrow in (a) indicate the vortex and anti-vortex centering
around the pair of nodes, which come to merge together in (c) and annihilate
into (d) under a strong AF field or decreasing doping. The
anti-vortex/vortex is not removed but driven into meron/anti-meron in (b).
Choice of parameters for the demonstration are (a) $m_{s}/t=0.2,\ \protect%
\delta =0.1$, (b) $m_{s}/t=0.55,\ \protect\delta =0.1$, (c) $m_{s}/t=0.2,\
\protect\delta =0.06$, (d) $m_{s}/t=0.2,\ \protect\delta =0.055.$}
\label{WeakPairingDwavePhaseTransitionSpcEvol}
\end{figure}

The other way of gapping out the nodes is to directly introduce a mass term
upon the pair of Weyl fermion-like quasiparticles. This can be achieved in
the process from the nodal $d$-wave SC to the $(s+id)_{w}$ SC with the
emergent additional extended $s$-wave pairing component. The corresponding
low-energy effective Hamiltonian Eq.(\ref{NodalEffHam}) is then changed into
\begin{eqnarray}
H_{\text{eff}}\left( \mathbf{K}_{\pm }+\mathbf{q}\right)  &=&\pm
v_{3}q_{+}\rho _{z}+v_{1}q_{-}\rho _{x}\mp (4\Delta _{s}\text{cos}K_{+})\rho
_{y}  \notag \\
&\equiv &\vec{h}_{\pm }(\mathbf{q})\cdot \vec{\rho}.
\end{eqnarray}%
The corresponding dispersion evolves as shown in Fig.~\ref%
{WeakPairingDwavePhaseTransitionSpcEvol}. It is important to point out that
the mass term introduced via $s_{x^{2}+y^{2}}$ pairing does not suppress the
topology of the nodal $d$-wave SC. At the cost of breaking the time-reversal
symmetry $\mathcal{T}$, the $s_{x^{2}+y^{2}}$ pairing contributes to the
out-of-plane component of $\vec{h}_{\pm }(\mathbf{k})$, driving the
anti-vortex/vortex texture into a meron/anti-meron instead, forming a
skyrmion. The skyrmion is approximately localized in the valleys, and is
complete only in the low-energy limit, where it loses sight of the Brillouin
zone and is absolutely isolated from its mirror partner (the anti-skyrmion
living on the other valley). Consequently, the $(s+id)_{w}$ SC in essence
realizes the topological crystalline SC. Formally, the topological Chern
number can be calculated and the weak to strong pairing transition of ($s$+i$%
d$) SC should be expected.

\section{Weak to strong $s+id$ pairing SC}

The ($s$+i$d$) SC phase is divided into two fully gapped $(s+id)_{w}$ and $%
(s+id)_{s}$ SC by a critical line. The critical line is characterized by the
nodes on the crossing point of the $d_{x^{2}-y^{2}}$ and $s_{x^{2}+y^{2}}$
nodal lines, i.e. $\mathbf{X}_{\pm }\equiv (\frac{\pi }{2},\pm \frac{\pi }{2}%
)$. Near the critical point the low-energy effective Hamiltonian can be
obtained by expanding the MF Hamiltonian for $\Psi _{-}(\mathbf{k})$ around
the valleys $\mathbf{X}_{\pm }$ to leading order, because $\Psi _{+}(\mathbf{%
k})$ is frozen in the low-energy limit. Then we have
\begin{eqnarray}
&&\widehat{H}_{eff}(\mathbf{X}_{\pm }+\mathbf{q})\equiv \vec{n}_{\pm }(%
\mathbf{q})\cdot \vec{\rho}  \label{critical} \\
&=&(-\mu ^{\prime }-Aq_{\pm }^{2}-A^{\prime }q_{\mp }^{2})\rho _{z}+2{q}%
_{\mp }\Delta _{d}\rho _{x}+2{q}_{\pm }\Delta _{s}\rho _{y},  \notag
\end{eqnarray}%
where $\mu ^{\prime }=\mu +m_{s}$, $A=2(t\delta +\chi )^{2}/m_{s}-t^{\prime
}\delta $, $A^{\prime }=t^{\prime }\delta $, ${q}_{+}=q_{x}+q_{y}$ and $%
q_{-}=q_{x}-q_{y}$. Within our phase diagram $A>0$ and $A^{\prime }>0$, the
effective chemical potential $\mu ^{\prime }$ controls the phase transition:
$\mu ^{\prime }<0$ accounts for $(s+id)_{w}$ SC phase while $\mu ^{\prime }>0
$ represents $(s+id)_{s}$ SC phase. $\mu ^{\prime }=0$ gives rise to the
critical point. Note that in phenomenological sense Eq.(\ref{critical})
resembles two copies of the effective Hamiltonian of weak pairing $(p_{x}\pm
ip_{y})$ SC discussed by Read and Green\cite{ReadGreen}, analogous to the
time-reversal invariant topological superconductor\cite{XLQ}, where $%
(p_{x}+ip_{y})_{\uparrow \uparrow }$ and $(p_{x}-ip_{y})_{\downarrow
\downarrow }$ are related by $\mathcal{T}$. However, $\widehat{H}_{eff}(%
\mathbf{X}_{+}+\mathbf{q})$ and $\widehat{H}_{eff}(\mathbf{X}_{-}+\mathbf{q})
$ living on the two valleys $\mathbf{X}_{\pm }$ are related by mirror
symmetries $M_{x/y}$ instead of the symmetry $\mathcal{T}$. Nevertheless,
our discussion of its topology goes quite parallel.

In the fully gapped $s$+i$d$ pairing states, since both $\Delta _{s,d}\neq 0$%
, the pairing gap function $\Delta _{\mathbf{k}}$ is complex in Eq.(\ref%
{critical}), where $\vec{n}(\mathbf{k})$ pins down the Anderson's pseudospin%
\cite{AndersonPseudospin}, giving rise to the SC ground state wave function
as
\begin{equation}
\left\vert \Omega \right\rangle \propto \text{exp}(\sum_{\mathbf{k}}g_{%
\mathbf{k}}b_{\mathbf{k}\uparrow }b_{-\mathbf{k}\downarrow })\left\vert
\text{FS}\right\rangle ,
\end{equation}%
where $g_{\mathbf{k}}=(1+\hat{n}_{z})/(\hat{n}_{x}-i\hat{n}_{y})$ with $\hat{%
n}=\vec{n}/|\vec{n}|$. Note that the SC ground state is written in the hole
representation, because the normal state Fermi sea in the low-energy theory
is given by the hole band $\xi _{-}(\mathbf{k})$, leading to the hole pocket
Fermi surfaces around $X_{\pm }$ for $\mu ^{\prime }<0$. It is then
straightforward to show that
\begin{equation}
g_{X_{\pm }}\propto \frac{1}{(\Delta _{s}q_{\pm }+i\Delta _{d}q_{\mp })},\
(\mu ^{\prime }<0)
\end{equation}%
which signifies singularity on $\mathbf{X}_{\pm }$ for $(s+id)_{w}$ pairing
phase, resulting in a modulated long tail of the pairing wave function in
real space. Actually, the function $g_{X_{\pm }}(\mathbf{k})$ defines a map
from the $\mathbf{k}$-space torus $T^{2}$ to a sphere $S^{2}$ parametrized
by the corresponding pseudo-spinor. Such a map is classified by the homotopy
group $\pi _{2}(S^{2})$, and the singular vortex structure of $g_{X_{\pm }}$
is identical to a topological monopole charge\cite{ReadGreen}. As shown
schematically in Fig.\ref{edgeSpc}a, the singular vortices residing on the
two valleys $\mathbf{X}_{\pm }$ are mirror partners in $(s+id)_{w}$ phase,
so that they carry monopoles with opposite topological charges $Q_{X_{\pm
}}=\pm 1$. In contrast, in the $(s+id)_{s}$ phase with $\mu ^{\prime }>0$,
the pairing wave function $g_{X_{\pm }}$ is analytic at $X_{\pm }$ and the
topological charge is zero, indicating that the strong pairing ($s$+i$d$)$%
_{s}$ phase corresponds to a topologically trivial phase.

From another perspective, since the pseudo-spinor is polarized by the unit
vector $\hat{n}(\mathbf{k})$, the $S^{2}$ of the topological map can be
alternatively spanned by $\hat{n}$, which naturally entails the topological
invariant Chern number $C=\frac{1}{4\pi }\int_{BZ}d^{2}k\ \hat{\mathbf{n}}%
\cdot (\partial _{k_{x}}\hat{\mathbf{n}}\times \partial _{k_{y}}\hat{\mathbf{%
n}})$ that characterizes the map. Physically, the Chern number is the
integral of the Berry curvature which counts the total Berry flux. Since the
Berry curvature is sharply peaked around the valley points $X_{\pm }$, the
total Berry flux can be approximately attributed as the sum of the Berry
flux carried by each valley $C=C_{X_{+}}+C_{X_{-}}$, in which $C_{X_{\pm }}$
converges fast to quantized value in the low-energy limit. By treating Eq.(%
\ref{critical}) in infinite large space\cite{Buettiker}, our calculation
shows that
\begin{eqnarray}
C_{X_{\pm }} &=&\frac{1}{4\pi }\int_{\infty }d^{2}q\ \hat{\mathbf{n}}_{%
\mathbf{X}_{\pm }}\cdot (\partial _{q_{x}}\hat{\mathbf{n}}_{\mathbf{X}_{\pm
}}\times \partial _{q_{y}}\hat{\mathbf{n}}_{\mathbf{X}_{\pm }})  \notag \\
&=&%
\begin{cases}
\pm 1, & \mu ^{\prime }<0, \\
0, & \mu ^{\prime }>0.%
\end{cases}%
.
\end{eqnarray}%
Although the total Chern number $C=0$ in both phases, it is meaningful to
introduce a valley Chern number\cite{Niu,Ezawa}, $C_{v}=C_{X_{+}}-C_{X_{-}}$
that describes the two topologically distinct phases and classifies the
topological phases by a Z$_{2}$ valley index. Note that the second order
terms in Eq.(\ref{critical}) are essential in removing the marginality,
giving rise to an integer instead of a half integer for $C_{X_{\pm }}$\cite%
{Volovik}. So the topological nontrivial phase and the trivial one are
distinguishable. To conclude, in low-energy limit we lose sight of the
Brillouin zone living in two inequivalent valleys $\mathbf{X}_{\pm }$, and
we have two copies of the nontrivial $(p_{x}\pm ip_{y})$-like topological SC
that are related by the mirror symmetry $M_{x/y}$ instead of time-reversal
symmetry $\mathcal{T}$ for the $(s+id)_{w}$ phase. This weak pairing SC is
characterized by a nonzero topological valley Chern number $C_{v}$ (Fig.~\ref%
{edgeSpc}a). In contrast, the $(s+id)_{s}$ SC is a trivial strong pairing
phase.

There's one thing remained to be addressed, i.e., the protecting symmetry of
the $(s+id))w$ SC phase. Since the two inequivalent valleys form a
valley-spinor, we denote the Pauli matrices acting on this spinor as $\gamma
_{\alpha =x,y,z}$, with the valley on $\mathbf{X}_{\pm }$ being the
eigen-spinor of $\gamma _{z}=\pm 1$, respectively. In this way, the
low-energy effective Hamiltonian Eq.(\ref{critical}) can be rewritten as
\begin{eqnarray}
\widehat{H}_{eff}(\mathbf{q}) &=&(-\mu ^{\prime }-Bq^{2})\rho
_{z}+2q_{x}(\Delta _{d}\rho _{x}+\Delta _{s}\rho _{y})  \notag \\
&&+(-B^{\prime }q_{x}\rho _{z}+\Delta _{s}\rho _{y}-\Delta _{d}\rho
_{x})2q_{y}\gamma _{z},
\end{eqnarray}%
where $B=A+A^{\prime }$ and $B^{\prime }=A-A^{\prime }$. The valley symmetry
$\mathcal{V}\equiv \gamma _{z}$ is preserved in the low-energy limit, and
the mirror reflection amounts to $\gamma _{x}$ that flips the two valleys.
As long as $\mathcal{V}$ is present, the valley-spinor is conserved and all
possible inter-valley couplings are forbidden, as a result the valley Chern
number $C_{v}$ is well-defined and cannot be changed without gap closing.
Namely, in the presence of valley symmetry $\mathcal{V}$, all mass terms for
$(s+id)_{c}$ are forbidden, so there is no way of adiabatically connecting
the $(s+id)_{w}$ and $(s+id)_{s}$ SC phases, evidencing their topological
distinction\cite{JackiwRebbi}. Analogous to the quantum spin Hall Hamiltonian%
\cite{Kane} and the $Z_{2}$ time-reversal invariant topological SC protected
by the time-reversal symmetry\cite{XLQ}, the $(s+id)_{w}$ SC phase is a
topological valley SC that nervertheless goes beyond the conventional
ten-fold way classification\cite{AZ,Schnyder,Kitaev}.

\section{Robust gapless edge states of $(s+id)_w$ SC}

\begin{figure}[t]
\includegraphics[width=8cm]{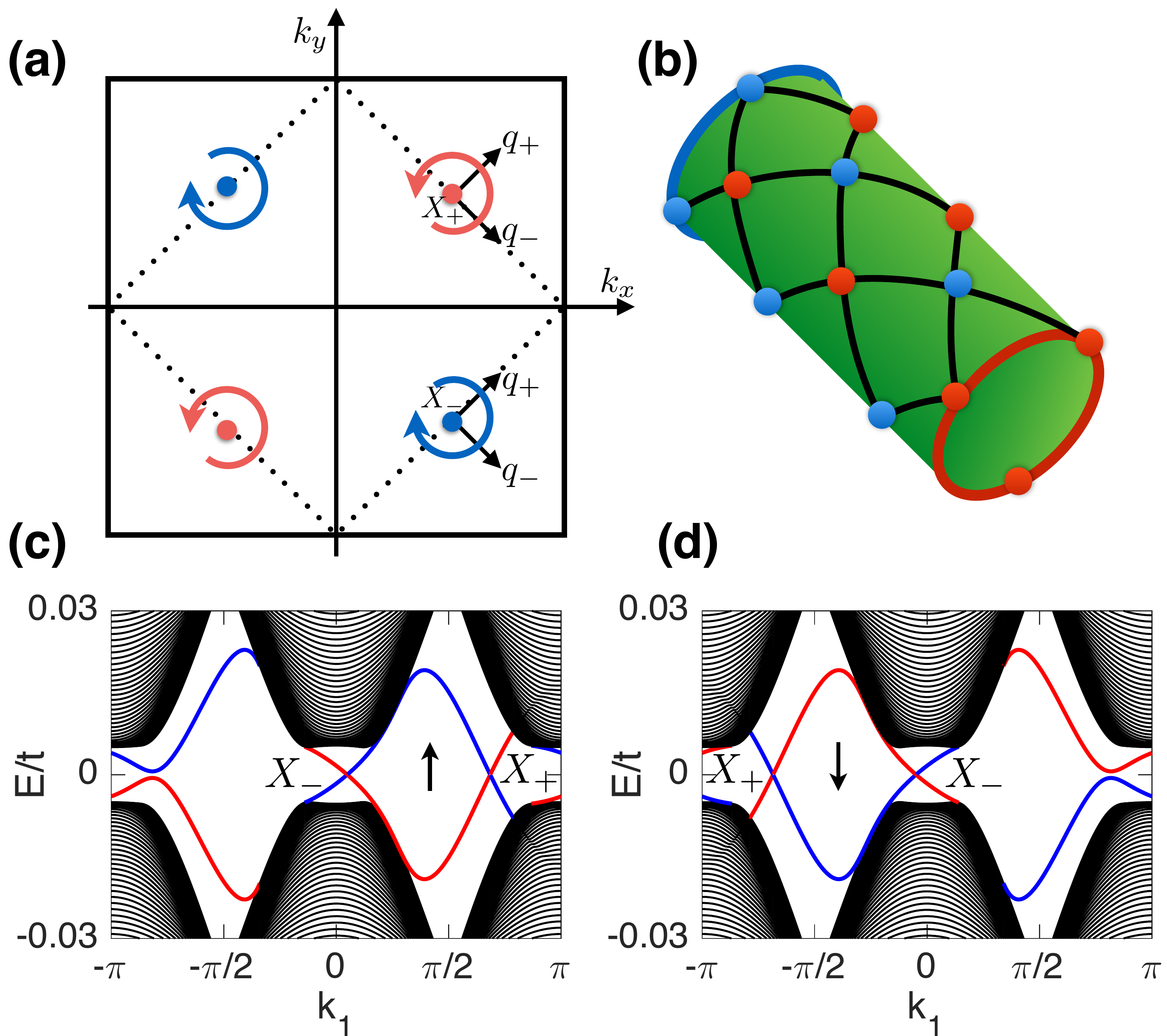}
\caption{(a) Low-energy effective theory resembles two copies of spinful $%
(p+ip)_{\uparrow \downarrow }$ and $(p-ip)_{\uparrow \downarrow }$ SC
components living in valleys around $\mathbf{X}_{+}$ and $\mathbf{X}_{-}$
respectively, which are related by mirror reflection symmetries $M_{x/y}$.
(b) Model in $(s+id)_{w}$ SC phase with $(1,\bar{1})$ edges is placed on a
finite size lattice with periodic boundary in $[1,1]$ direction i.e. of
cylinder geometry. The open edges are non-centrosymmetric and are denoted by
blue and red color, related by bond-centered mirror reflection $\tilde{M}%
_{x} $ which is the combination of site-centered mirror reflection and unit
lattice translation. (c) spin up and (d) spin down Bogoliubov quasi-particle
dispersion. Gray is for the bulk excitation while blue and red are for
quasiparticles localized on corresponding edges on the cylinder geometry
with $200\times 512$ lattice sites. ($\protect\delta =0.1$, $m_{s}/t=0.58$, $%
\Delta _{s}/t=0.0055$, $\Delta _{d}/t=0.0056$, $\protect\chi /t=0.0120$, $%
\protect\mu /t=-0.5861$). }
\label{edgeSpc}
\end{figure}

To demonstrate that the topological $(s+id)_{w}$ phase indeed supports
robust gapless edge modes, we perform exact diagonalization to the model
with $(1,\bar{1})$ open edges (Fig.\ref{edgeSpc}b). In the cylinder
geometry, the momentum $k_{1}$ along the edge remains a good quantum number
and the valley symmetry is preserved upon projection onto the boundary.
Within the surface Brillouin zone, the two valleys are located at $k_{1}=0,\
\pi $ respectively. For the singlet pairings, the Bogoliubov excitations are
spin polarized. The dispersions in the surface Brillouin zone are shown in
Fig.\ref{edgeSpc}c and \ref{edgeSpc}d for spin up and down Bogoliubov
quasi-particles, respectively, which are related by the particle-hole
transformation $\Xi =i\sigma _{x}\rho _{y}\mathcal{K}$. The spin index in
our BdG basis is chosen along the $y$ direction and the whole spectrum
satisfies particle-hole symmetry $\Xi ^{-1}E_{k}\Xi =-E_{-k}$.

Having a closer look at the spin up excitations, for instance, in the
vicinity of each valley there is one gapless chiral edge mode, and the edge
modes associated with the two valleys propagate in opposite directions, so
the whole system is non-chiral analogous to the quantum valley Hall effect.
This pattern concurs with the Chern number for each valley, because the
edges being the interface of the nontrivial bulk and the trivial vacuum
should close the gap and the number of edge modes is supposed to amount to
the mismatch of the Chern number between the bulk valley and the vacuum.
Since the Chern number for the valley on $\mathbf{X}_{\pm }$ is $\pm 1$,
there are supposed to be counter-propagating chiral edge modes associated
with the two valleys. However, it is worth noting that on the same edge the
edge currents contributed by the two valleys differ slightly by their
velocities, so that the net velocity is nonzero and can be detected
experimentally. This is the consequence of the noncentrosymmetry of the
lattice edges (Fig.~\ref{edgeSpc}b). In contrast, we do not observe any edge
modes on $(1,0)$ or $(0,1)$ edges, because the valleys would collapse onto
the same momentum on surface Brillouin zone and break valley symmetry.
Neither is there any sign of gapless edge modes in the strong-pairing $%
(s+id)_{s}$ phase, evidencing its trivial topology distinct from $(s+id)_{w}$%
.

\section{Discussion and conclusion}

Now that we have elaborated on the weak pairing topological nature of the
nodal $d$-wave SC and the $(s+id)_{w}$ SC as well as their phase transitions
into strong pairing trivial SC phases, it should be mentioned that the
properties of the nodal $s_{x^{2}+y^{2}}$-wave SC are similar to the nodal $%
d_{x^{2}-y^{2}}$-wave SC and the scenarios of its phase transition either
into nodeless $s_{x^{2}+y^{2}}$-wave phase or the $(s+id)_{w}$ phase are
parallel to that of the $d_{x^{2}-y^{2}}$-wave SC. To summarize the phase
diagram in Fig.~\ref{setup&phase_diagram}b, the nodal $d_{x^{2}-y^{2}}$%
-wave, $(s+id)_{w}$ and nodal $s_{x^{2}+y^{2}}$-wave are weak pairing and
topologically nontrivial. Among them the nodal phases are topological nodal
superconductor whose nodes carry topological number and entail edge modes on
edges residing between projection of nodes, and the fully gapped $(s+id)_{w}$
realizes topological valley SC. On the other hand, the nodeless $%
d_{x^{2}-y^{2}}$-wave and $(s+id)_{s}$ and nodeless $s_{x^{2}+y^{2}}$-wave
are all fully gapped strong pairing topologically trivial SC, between which
we find no gap closing phase transition. The phase transition from weak to
strong pairing phases necessarily experiences a critical point described by
effective Hamiltonian Eq.~\ref{critical}.

In summary, we have shown that when a nodal $d$-wave SC is proximately
coupled to an AF insulator, the nature of the SC can be remarkably changed
into nodeless $d$-wave, nodeless and nodal $s$-wave, and
valley-symmetry-protected $Z_{2}$ topological $(s+id)_{w}$ SC phases,
depending on the dopant concentration and the proximity induced AF field.
These findings are supported by careful studies of the SC phases described
by the two-dimensional t-J model. The presence of a AF field is crucial, and
its existence can be justified in the CuO$_{2}$ monolayer on the substrates
of optimal doped cuprates\cite{Xue}. Our theoretical calculations suggest
that a possible candidate for the nodeless superconductivity observed in the
CuO$_{2}$ monolayer on the optimally doped Bi-2212 substrates may be the
valley-symmetry-protected topological SC, consistent with the large overall
Mott gap found in the scanning tunneling spectrum. However, so far it is not
clear how strong the AF field in the CuO$_{2}$ insulating layer can be
reached experimentally. Our findings certainly broaden the scope of the
investigations on high Tc cuprates to include the possibility of topological
valley superconductivity. In order to confirm such novel nodeless SCs in the
cuprates, further new experiments are desirable to verify our predictions.

The authors are indebted to Kun Jiang for his helpful discussion, especially
on the identification of the protecting symmetry of the topological
superconducting phase. G.M.Z. acknowledges the support of National Key
Research and Development Program of China (2016YFA0300300) and NSF-China
through Grant No.20121302227. Z.W. are supported by the U.S. Department of
Energy, Basic Energy Sciences, under Award DE-FG02-99ER45747.

\newpage


\begin{thebibliography}{99}
\bibitem{Muller} J. P. Bednorz and K. A. Muller, Z. Phys. B \textbf{64}, 189
(1986).

\bibitem{Anderson} P. W. Anderson, Science 235, 1196 (1987).

\bibitem{AtoZ} P. W. Anderson, P. A. Lee, M. Randeria, T. M. Rice, N.
Trivedi, and F. C Zhang, J. Phys. Condens. Matter \textbf{16}, R755 (2004).

\bibitem{LeeNagaosaWen} P. A. Lee, N. Nagaosa, and X. G. Wen, Rev. Mod.
Phys. \textbf{78}, 17 (2006).

\bibitem{Shen} Z. X. Shen, D. S. Dessau, B. O. Wells, D. M. King, W. E.
Spicer, A. J. Arko, D. Marshall, L. W. Lombardo, A. Kapitulnik, P.
Dickinson, S. Doniach, J. DiCarlo, T. Loeser, and C. H. Park, Phys. Rev.
Lett. \textbf{70}, 1553 (1993).

\bibitem{Harlingen} D. A. Wollman, D. J. Van Harlingen, W. C. Lee, et al,
Phys Rev Lett. \textbf{71}, 2134 (1993)

\bibitem{Tsuei} C. C. Tsuei, J. R. Kirtley, C. C. Chi, L. S. Yujahnes, A.
Gutpa, T. Shaw, J. Z. Sun, and M. B. Ketchen, Phys. Rev. Lett. \textbf{73},
593 (1994).

\bibitem{Xue} Y. Zhong, Y. Wang, S. Han, Y. F. Lv, W. L. Wang, D. Zhang, H.
Ding, Yi. M. Zhang, L. L. Wang, K. He, R. D. Zhong, J. A. Schneeloch, G. D.
Gu, C. L. Song, X. C. Ma, Q. K. Xue, Science Bulletin \textbf{61}, 1239
(2016), arXiv:1607.01852.

\bibitem{ZhangRice} F. C. Zhang, and T. M. Rice, Phys. Rev. B \textbf{37},
3759 (1988).

\bibitem{KotliarLiu} G. Kotliar and J. Liu, Phys. Rev. B \textbf{38}, 5142
(1988).

\bibitem{Joynt} B. E. C. Koltenbah and R. Joynt, Rep. Prog. Phys. \textbf{60}%
, 23 (1997) and references therein.

\bibitem{ReadGreen} N. Read and Dmitry Green, Phys. Rev. B \textbf{61},
10267 (2000).

\bibitem{ZhuZhang} G. Y. Zhu and G. M. Zhang, arXiv: 1702.07090 (2017).

\bibitem{Kivelson} E. Berg, C-C. Chen, and S.A. Kivelson, Phys. Rev. Lett.
\textbf{100}, 027003 (2008).

\bibitem{Yuanming} Y.-M. Lu and Z. Wang, Phys. Rev. Lett. \textbf{110},
096403 (2013).

\bibitem{XLQ} X. L. Qi, T. L. Hughes, S. Raghu, S. C. Zhang, Phys. Rev.
Lett. \textbf{102}, 187001 (2009).

\bibitem{AndersonPseudospin} P.W. Anderson, Phys. Rev. \textbf{110}, 827;
\textbf{112}, 1900 (1958).

\bibitem{Buettiker} J. Li, A. F. Morpurgo, M. Buettiker, and I. Martin,
Phys. Rev. B \textbf{82}, 245404 (2010).

\bibitem{Niu} D. Xiao, W. Yao, and Q. Niu, Phys. Rev. Lett. \textbf{99},
236809 (2007).

\bibitem{Ezawa} Motohiko Ezawa, Physics Letters A \textbf{378}, 1180 (2014).

\bibitem{Volovik} G. E. Volovik, The Universe in a Helium Droplet (Oxford
University Press, Oxford) (2003).

\bibitem{JackiwRebbi} R. Jackiw and C. Rebbi, Phys. Rev. D \textbf{13}, 3398
(1976).

\bibitem{Kane} C.L. Kane, E.J. Mele, Phys. Rev. Lett. \textbf{95}, 226801
(2005).

\bibitem{AZ} A. Altland and M.R. Zirnbauer, Phys. Rev. B \textbf{55}%
,1142-1161 (1997).

\bibitem{Schnyder} A. P. Schnyder, S. Ryu, A. Furusaki and A. W. W. Ludwig,
Phys. Rev. B \textbf{78}, 195125 (2008).

\bibitem{Kitaev} Kitaev, AIP Conf. Proc. \textbf{1134}, 22-30 (2009).
\end{thebibliography}
\end{document}